\documentclass[12pt]{article}
\usepackage[centertags]{amsmath}
\usepackage{amsfonts}
\usepackage{amssymb}
\usepackage{graphicx}
\usepackage{amssymb}
\usepackage[verbose,colorlinks=true,naturalnames=true,linkcolor=blue,]
{hyperref}
\newcommand{\sign}{\mathop{\rm sign}\nolimits}
\newcounter{rown}

\begin{document}

\title{Jordanian  Quantum Deformations of \\
\vskip 5pt
$D=4$ Anti-de-Sitter and Poincar\'{e} Superalgebras}
\author{A. Borowiec$^{1)}$, J. Lukierski$^{1)}$ and V.N. Tolstoy$^{1),2)}$
\\ \\
$^{1)}$Institute for Theoretical Physics,
\\University of Wroc{\l}aw, pl. Maxa Borna 9,
\\50--205 Wroc{\l}aw, Poland\\
\\$^{2)}$Institute of Nuclear Physics,
\\Moscow State University, 119 992 Moscow, Russia}

\date{\empty}
\maketitle

\begin{abstract}
We consider a superextension of the extended Jordanian twist,
describing nonstandard quantization of anti-de-Sitter ($AdS$)
superalgebra $\mathfrak{osp}(1|4)$ in the form of Hopf superalgebra.
The super-Jordanian twisting function and corresponding basic
coproduct formulae for the generators of $\mathfrak{osp}(1|4)$ are
given in explicit form. The nonlinear transformation of the
classical superalgebra basis not modifying the defining algebraic
relations but simplifying coproducts and antipodes is proposed. Our
physical application is to interpret the new super-Jordanian
deformation of $\mathfrak{osp}(1|4)$ superalgebra as deformed $D=4$
$AdS$ supersymmetries. Subsequently we perform suitable contraction
of quantum Jordanian $AdS$ superalgebra and obtain new
$\kappa$-deformation of $D=4$ Poincar\'{e} superalgebra, with the
bosonic sector describing the light cone $\kappa$-deformation of
Poincar\'{e} symmetries.
\end{abstract}

\section{Introduction}
Main aims of studying the quantum deformations of space-time Lie
algebras and Lie superalgebras is to provide the geometric origin
of deformed relativistic symmetries, noncommutative space-times
and their corresponding supersymmetric extensions. The quantum
deformations of Poincar\'{e}, $AdS$ and conformal space-time
symmetries in  $D=4$  were already  extensively studied (see e.g.
\cite{LNRT} -- \cite{LLM2}). In particular for the Poincar\'{e}
and conformal algebras one can introduce the deformations
parametrized by a geometric mass parameter $\kappa$, i.e.
described as so--called $\kappa$-deformations (see e.g.
\cite{LNRT,MR,LRZ,LMM}--\cite{LLM2}) which were also used
for the introduction of $\kappa-$deformed field theories
(see e.g. \cite{KoLM2}--\cite{M4}).
Such deformations introduce
in geometric way third fundamental parameter $\kappa$ in physics
(besides $\hbar$ and $c$) which can be linked with Planck mass and
quantum gravity \cite{DFR1,ASS}. It appears that in general case, e.g. for $D=4$
conformal algebra, one can introduce several mass-like deformation
parameters \cite{LLM1}.

It is well-known that in recent quarter of the last century the new
unified models of fundamental interactions are supersymmetric (e.g
supergravities, superstrings, super-$p$-branes, super-$D$-branes,
$M$-theory). In particular, we stress that the noncommutative
space-times describing $D$-brane world-volume coordinates in Kalb-Ramond
two-tensor background \cite{SWitten} in fact should be extended
supersymmetrically (see e.g. \cite{FL}--\cite{Sei}). We argue therefore  that if
the notion of noncommutative geometry and quantum groups are
applicable to the present supersymmetric framework of 
fundamental interactions, the noncommutative superspaces as well as the quantum
supersymmetries should be studied. In this paper we limit ourselves
to the case of "physical" $D=4$ SUSY case; the application to e.g.
$M-$theory requires the consideration of deformed supersymmetries
and superspaces for all $D\le 11$.

For $D=4$ supersymmetries only the standard $\kappa$-deformation
of the Poincar\'{e} superalgebra has been studied
\cite{LNR}--\cite{KoLM} which was obtained as the quantum $AdS$
contraction of the Drinfeld-Jimbo $q$-deformation of
$\mathfrak{osp}(1|4)$, in the contraction limit
$\lim\limits_{{q\to1}\atop{R\to\infty}}R\ln q\,=\,\kappa^{-1}$, where
$R$ is the $AdS$ radius. In such a limit  the classical $r$-matrix
\begin{equation}\label{v1}
r\;=\;\frac{1}{\kappa}\,\sum_{i=1}^3 N_i\,\wedge\,P_i
\end{equation}
describing  standard $\kappa$-deformed $D=4$ Poincar\'{e} algebra
was supersymmetrized. We recall that the elements $M_j =
\frac{1}{2} \epsilon_{jkl}{}M_{kl}$, $N_j = M_{0j}$, $P_k$, $P_0$
$(j=1,2,3)$ generating  the Poincar\'{e} algebra
${\cal{P}}(3,1)=\{M_{\mu\nu},P_{\mu}\,|\,\mu,\nu=0,\,\ldots,4\}$
satisfy the standard commutation relations
\begin{equation}
\begin{array}{rcccccl}\label{v2}
[M_j,\,M_k]\!\!&=\!\!& i\epsilon_{jkl}\,M_l~,\quad\;
[M_j,\,N_k]\!\!&=\!\!&i\epsilon_{jkl}\,N_l~,\quad
[N_j,\,N_k]\!\!&=\!\!&-i\epsilon_{jkl}\,M_l,
\\[5pt]
[M_j,\,P_k]\!\!&=\!\!& i\epsilon_{jkl}\,P_l~,\qquad\!
[M_j,\,P_0]\!\!&=\!\!&0~,\phantom{aaaaaaaaaaaa}&&
\\[5pt]
[N_j,\,P_k]\!\!&=\!\!&-i\delta_{jk}\,P_0~,\quad\;
[N_j,\,P_0]\!\!&=\!\!&-iP_j^{}~,\qquad
[P_\mu,\,P_\nu]\!\!&=\!\!&0~.
\end{array}
\end{equation}

In this paper we consider other nonstandard deformation of
$\mathfrak{osp}(1|4)$ superalgebra with the classical $r$-matrix
of Jordanian type, satisfying graded classical Yang-Baxter
equation \cite{T3,CK1}. We show that in the quantum contraction
limit $R\to\infty$ the fundamental mass parameter $\kappa$
entering into our new $\kappa$-deformed $D=4$ super-Poincar\'{e}
algebra is defined as follows
\begin{equation}\label{v3}
\lim\limits_{{\xi\to0}\atop{R\to\infty}}\xi\,R\,=\frac{i}{\kappa}~.
\end{equation}
where $\xi$ denotes suitably chosen dimensionless parameter
describing the nonstandard deformation of $\mathfrak{osp}(1|4)$.
In such a way we obtain an alternative $\kappa$-deformation of
$D=4$ super-Poincar\'{e} algebra described infinitesimally by the
supersymmetrization of the following $r$-matrix for light-cone
$\kappa$-deformation of $D=4$ Poincar\'{e} algebra
\cite{LMM,LLM1}:
\begin{equation}\label{v4}
r\,=\,\frac{1}{\kappa}\Bigl(P_1^{}\wedge (N_1^{}+M_2^{})
+P_2{}\wedge (N_2^{}-M_1^{})+P_+\wedge N_3^{}\Bigr)~,
\end{equation}
where $P_{+}=P_0+P_3$. The quantum deformation of relativistic
symmetries generated by the $r$-matrix (\ref{v4}) describes the
$\kappa$-deformed Minkowski space with the "quantized" light-cone
direction $x_+=x_0+x_3$ \cite{KoM,BHOS}. We point out that the
light-cone $\kappa$-Poincar\'{e} algebra has been introduced
firstly in a particular basis under the name of the null-plane
quantum Poincar\'{e} algebra \cite{BHOS}.

In a general case the classical $r$-matrices of Jordanian type for
any simple Lie algebra depend on several deformation parameters,
$\xi_1,\;\xi_2,\ldots,\xi_n$, and it is a sum of one-parameter
classical $r$-matrices of Jordanian type \cite{T3,KLM,LO},
where the one-parameter classical $r$-matrix has the form
\begin{equation}
r(\xi)=\xi\,\Bigl(h_{\gamma_0^{}}^{}\wedge e_{\gamma_0^{}}^{}+
\sum_{i=1}^{N}e_{\gamma_i}^{}\wedge e_{\gamma_{-i}}^{}\Bigr)~,
\label{v5}
\end{equation}
provided that the generators $(h_{\gamma_0}^{}$,
$e_{\gamma_0}^{}$, $e_{\gamma_{\pm i}^{}}$) satisfy the relations
(see \cite{T3})\footnote{The formulae (\ref{v5}) and (\ref{v6})
generalize the considerations presented in \cite{KLM,LO} by
describing the classical $r$-matrices with the
support $\{h_{\gamma_0^{}},e_{\gamma_0^{}},e_{\gamma_{\pm
i}^{}}\}$ which does not necessarily belong to Borel subalgebra.}
\begin{equation}
\begin{array}{rcl}
[h_{\gamma_0^{}},\,e_{\gamma_0^{}}]\!\!&=\!\!&e_{\gamma_0^{}}~,
\qquad\qquad\;\,[e_{\gamma_{i}^{}},\,e_{\gamma_{-j}}]\;=\;
\delta_{ij}^{}\,e_{\gamma_0^{}}~,
\\[5pt]
[h_{\gamma_{0}^{}},\,e_{\gamma_i^{}}]\!\!&=&\!\!
(1-t_{\gamma_i^{}}^{})\,e_{\gamma_i^{}}~,\quad
[h_{\gamma_0^{}},\,e_{\gamma_{-i}^{}}]\;=\;
t_{\gamma_i^{}}^{}\,e_{\gamma_{-i}^{}}
\quad (t_{\gamma_i^{}}^{}\in{\mathbb C})~,
\\[5pt]
[e_{\gamma_{\pm i}^{}},\,e_{\gamma_0^{}}^{}]\!\!&=\!\!&0~,\qquad
\qquad\;\;\,[e_{\gamma_{\pm i}^{}},\,e_{\gamma_{\pm j}}]\;=\;0~.
\label{v6}
\end{array}
\end{equation}

In the case of Lie superalgebras the terms constructed from odd
(fermionic) generators should be added in (\ref{v5}) (see
\cite{T3}). For the Lie superalgebra $\mathfrak{osp}(1|4)$ the
general classical $r$-matrix of Jordanian type is two-parameter,
$r_{Ad S}(\xi_1,\xi_2)$. It turns out that the $AdS$ contraction
limit (\ref{v2}) gives a non-trivial result provided
$\xi=\xi_1=\xi_2$, and it describes our new $\kappa$-deformation
of $D=4$ super-Poincar\'{e} algebra.

Recently in \cite{BLT} by completing earlier results from
\cite{CK2} (see also \cite{ACS}) there was obtained a twist
quantization of $\mathfrak{osp}(1|2)$, describing the
$\kappa$-deformation of $D=1$ conformal superalgebra, which can be
interpreted as the deformation of $D=2$ $AdS$ superalgebra.
Similarly, in this paper we complete the results presented in
\cite{CK1} and consider the twist quantization of
$\mathfrak{osp}(1|4)$ with the physical application to $AdS$
supersymmetry and its super-Poincar\'{e} limit.

The plan of our paper is the following. In Sec.~2 we present the
mathematical (Cartan-Weyl) basis of $\mathfrak{osp}(1|4)$ and
consider two corresponding Jordanian type classical $r$-matrices.
In Sec.~3 we present the two-parameter quantization of these
classical $r$-matrices with deformation modifying only coalgebra
sector, and calculate basic coproducts for $\mathfrak{osp}(1|4)$
generators. Further, employing the formulas recently proposed in
\cite{T3} we introduce new more suitable basis for the
superalgebra $\mathfrak{osp}(1|4)$. In Sec. 4 we shall introduce
the $AdS$ physical basis and perform the $AdS$ contraction
introducing mass-like deformation parameter $\kappa$. In such a
limit we obtain the Jordanian type classical $r$-matrices and the
twisting two-tensors for Poincar\'{e} and super-Poincar\'{e}
algebras. In Sec.~5 we comment on $\kappa$-deformations of $N=1$
Poincar\'{e} supersymmetries and on deformed $N$--extended $AdS$
supersymmetries.

\section{Cartan-Weyl basis and Jordanian type classical\\
$r$-matrices of $\mathfrak{osp}(1|4)$}
\setcounter{equation}{0}

In order to obtain compact formulas describing the commutation
relations for generators of the orthosymplectic superalgebra
$\mathfrak{osp}(1|4)$ we use embedding of this superalgebra in the
general linear superalgebra $\mathfrak{gl}(1|4)$. For convenience
we consider the general case of $\mathfrak{osp}(1|2n)$ \cite{T1}
embedded in $\mathfrak{gl}(1|2n)$. Let $a_{ij}^{}$ ($i,j=0,\pm1,$
$\pm2, \ldots,\pm n$) be a standard basis\footnote{This basis can
be realized by graded $(2n+1)\times(2n+1)$-matrices.} of the
superalgebra $\mathfrak{gl}(1|2n)$ with the standard
supercommutation relations
\begin{equation}\label{cw1}
[a_{ij}^{},a_{kl}^{}]:=a_{ij}^{}
a_{kl}^{}-(-1)^{\theta_{ij}\theta_{kl}}a_{kl}^{}
a_{ij}^{}=\delta_{jk}^{}a_{il}^{}-
(-1)^{\theta_{ij}\theta_{kl}}\delta_{il}^{}a_{kj}^{}~,
\end{equation}
where $\theta_{ij}^{}=1$ when one index $i$ or $j$ is equal to 0
and another takes any value $\pm1,\ldots,\pm n$;
 $\theta_{ij}^{}=0$ in the remaining  cases. The superalgebra
$\mathfrak{osp}(1|2n)$ is embedded in $\mathfrak{gl}(1|2n)$ as a
linear envelope of the following generators:
\\(i) the even (boson) generators spanning  the symplectic algebra
$\mathfrak{sp}(2n)$:
\begin{equation}\label{cw2}
e_{ij}^{}:=a_{i-j}^{}+\sign(ij)\,a_{j-i}^{}=
\sign(ij)\,e_{ji}^{}\quad(i,j=\pm1,\pm2,\ldots,\pm n)~;
\end{equation}
(ii) the odd (fermion) generators extending $\mathfrak{sp}(2n)$
 to $\mathfrak{osp}(1|2n)$:
\begin{equation}\label{cw3}
e_{0i}^{}:=a_{0-i}^{}+\sign(i)\,a_{i0}^{}=
\sign(i)\,e_{i0}^{}\quad(i=\pm1,\pm2,\ldots,\pm n)~.
\end{equation}
We also set $e_{00}^{}=0$ and introduce the sign function: $\sign
x=1$ if a real number $x\geq0$ and $\sign x=-1$ if $x<0$. One can
check  that the elements (\ref{cw2}) and (\ref{cw3}) satisfy the
following relations:
\begin{eqnarray}\label{cw4}
[e_{ij}^{},\,e_{kl}^{}]\!\!&=\!\!&
\delta_{j-k}^{}e_{il}^{}+\delta_{j-l}^{}
\sign(kl)\,e_{ik}^{}-\delta_{i-l}^{}e_{kj}^{}-
\delta_{i-k}^{}\sign(kl)\,e_{lj}^{} ~,
\\[3pt]
[e_{ij}^{},\,e_{0k}^{}]\!\!&=\!\!&
\delta_{j-k}^{}\sign(k)\,e_{i0}^{}-
\delta_{i-k}^{}e_{0j}^{}~,\label{cw5}
\\[3pt]
\{e_{0i}^{},e_{0k}^{}\}\!\!\!&=\!\!& \sign(i)\,e_{ik}^{}
\label{cw6}
\end{eqnarray}
for all $i,j,k,l=\pm1,\pm2,\ldots,\pm n$, where the bracket
$\{\cdot,\cdot\}$ means anti-commutator.

In our case of $\mathfrak{osp}(1|4)$ we have $n=2$. The $24$
elements $e_{ij}^{}$ $(i,j=0,\pm1,\pm2)$ are not linearly
independent (we have 10 constraints, for example,
$e_{1-2}^{}=-e_{-21}^{})$ and we can choose from them the
Cartan-Weyl basis as follows
\begin{eqnarray}\label{cw7}
&& (a)\; {\rm the\; rising\; generators}:\;\;
e_{1-2}^{},\,\,\,e_{12}^{},\,\,\,e_{11}^{},\,\,\,e_{22}^{},
\,\,\,e_{01}^{},\,\,\,e_{02}^{}~;
\\[3pt]\label{cw8}
&& (b)\; {\rm the\; lowering\; generators}:\;\;
e_{2-1}^{},\,\,\,e_{-2-1}^{},\,\,\,e_{-1-1}^{},\,\,\,e_{-2-2}^{},
\,\,\,e_{-10}^{},\,\,\,e_{-20}^{}~;
\\[3pt]\label{cw9}
&& (c)\; {\rm the\; Cartan\; generators}:\;\; h_{1}^{}:=
e_{1-1}^{},\,\,\,h_{2}^{}:= e_{2-2}^{}~.
\end{eqnarray}
In accordance with \cite{T3}, \cite{CK1} the general formula for
the Jordanian type classical $r$-matrix of  $\mathfrak{osp}(1|4)$
is given as follows
\begin{eqnarray}
r_{1,2}^{}(\xi_1^{},\xi_{2}^{})\,=\,r_{1}^{}(\xi_1^{})+
r_{2}^{}(\xi_2^{})~, \label{cw10}
\end{eqnarray}
where the classical $r$-matrices of Jordanian type
$r_{1}^{}(\xi_1^{})$ and $r_{2}^{}(\xi_2^{})$ have the form
\begin{eqnarray}
r_{1}^{}(\xi_1^{})\!\!&=\!\!&\xi_1^{}\Bigl(\frac{1}{2}\,e_{1-1}^{}
\wedge e_{11}^{}+e_{1-2}\wedge e_{12}-2e_{01}^{}\otimes e_{01}^{}
\Bigr)~,
\label{cw11}
\\[5pt]
r_{2}^{}(\xi_2^{})\!\!&=\!\!&\xi_2^{}\Bigl(\frac{1}{2}\,e_{2-2}^{}
\wedge e_{22}^{}-2e_{02}^{}\otimes e_{02}^{}\Bigr)~. \label{cw12}
\end{eqnarray}
Below we shall describe the twist quantization generated by the
classical $r$-matrix (\ref{cw10}).

\section{Jordanian type deformations of $\mathfrak{osp}(1|4)$}
\setcounter{equation}{0}

In accordance with the general scheme (see  \cite{T3}) the
complete twisting two-tensor $F(\xi_1,\xi_2)$ corresponding to the
resulting Jordanian type $r$-matrix (\ref{cw10}) is given as
follows
\begin{equation}
F(\xi_1^{},\xi_2^{})\;=\;\tilde{F}_{22}^{}(\xi_2^{})
F_{11}^{}(\xi_1^{})~, \label{jt1}
\end{equation}
where $F_{11}^{}(\xi_1^{})$ is the twisting two-tensor
corresponding to the classical $r$-matrix (\ref{cw11}), and
$\tilde{F}_{22}^{}(\xi_2^{})$ is the transformed twisting
two-tensor corresponding to the classical $r$-matrix (\ref{cw12})
(see below the formulas (\ref{jt15})). Thus we can implement full
deformation in two steps corresponding to two terms in the formula
(\ref{cw10}):

\underline{\it The first step of Jordanian type deformation}. The
twisting two-tensor $F_{11}^{}(\xi_1^{})$ corresponding to the
$r$-matrix (\ref{cw11}) has the form
\begin{equation}
F_{11}^{}(\xi_1^{})\;=\;\mathfrak{F}_{11}^{}(\xi_1^{})
F_{\sigma_{11}^{}}
\label{jt2}
\end{equation}
where the two-tensor $F_{\sigma_{11}^{}}$ is the Jordanian twist
\cite{O} and $\mathfrak{F}_{11}^{}$ is the super-extension of the
Jordanian twist. These two-tensors are given by the formulas
\begin{equation}
F_{\sigma_{11}^{}}\;= \;e^{e_{1-1}^{}\!\otimes\sigma_{11}^{}}~,
\label{jt3}
\end{equation}
\begin{equation}
\mathfrak{F}_{11}^{}(\xi_1^{})\,=\,\exp\big(\xi_1^{}
e_{1-2}^{}\otimes e_{12}^{}e^{-\sigma_{11}^{}}\big)\,
\mathfrak{F}_{01}^{}~,
\label{jt4}
\end{equation}
where the first factor on rhs of (\ref{jt4}) describes extended
Jordanian twist \cite{KLM}, and \cite{BLT}
\begin{eqnarray}
\mathfrak{F}_{01}^{}\!\!&=\!\!&\sqrt{\frac{(1+e^{\sigma_{11}^{}})\otimes
(1+e^{\sigma_{11}^{}})}{2(1+e^{\sigma_{11}^{}}\!\otimes
e^{\sigma_{11}^{}})}}\;\Bigl(1-2\xi_1^{}\frac{e_{01}^{}}
{1+e^{\sigma_{11}^{}}}\otimes\frac{e_{01}^{}}
{1+e^{\sigma_{11}^{}}}\Bigr)~,
\\[5pt]
\sigma_{11}^{}\!\!&=\!\!&\frac{1}{2}\ln(1+\xi_1^{}e_{11}^{})~.
\label{jt6}
\end{eqnarray}
We shall not provide here the explicit formulas of the twisted
coproduct $\Delta_{\xi_1^{}}(x):=F_{11}^{}(\xi_1^{})\Delta(x)
F_{11}^{-1}(\xi_{1}^{})$ and the corresponding antipode
$S_{\xi_1^{}}(x)$ for the generators $x=e_{ik}$ since these
formulas are intermediate in our scheme\footnote{The coproduct
formulas for the generators spanning the classical $r$-matrix
(\ref{cw11}) can be found in \cite{T3}.}.

Let us introduce the new generators in $U(\mathfrak{osp}(1|4))$ by
the inner automorphism firstly proposed in \cite{T3}
\begin{eqnarray}
w_{\xi_1^{}}^{}:=\sqrt{u(\mathfrak{F}_{11}^{}(\xi_1^{}))}\!\!&=\!\!&
\exp\Big(\frac{\xi_1^{}\,\sigma_{11}^{}\,e_{1-2}^{}\,e_{12}^{}}
{1-e^{2\sigma_{11}^{}}}\Big)\,\exp\Bigl(\frac{1}{4}\,
\sigma_{11}\Bigr)~,
\label{jt7}
\end{eqnarray}
where $u(\mathfrak{F}_{11}^{}(\xi_{1}^{}))$ is the Hopf "folding"
of the two-tensor (\ref{jt4}):
$u(\mathfrak{F}_{11}^{}(\xi_1^{}))=((S_{\xi_1^{}}^{}\!\otimes\,
\mathop{\rm Id})\mathfrak{F}_{11}^{}(\xi_1^{}))\circ1$. We
postulate the similarity map
$\tilde{e}_{ik}^{}:=w_{\xi_1^{}}e_{ik}^{}w_{\xi_1^{}}^{-1}$
preserving the defining relations (\ref{cw4})--(\ref{cw6}).

The twisted coproduct for the elements $\tilde{e}_{ik}$ are given
by simple and uniform formulas:
\begin{eqnarray}
\Delta_{\xi_1^{}}(e^{\pm\tilde{\sigma}_{11}^{}})\!\!&=\!\!&
e^{\pm\tilde{\sigma}_{11}^{}}\otimes
e^{\pm\tilde{\sigma}_{11}^{}}~,
\\[7pt]
\Delta_{\xi_1^{}}(\tilde{e}_{ik}^{})\!\!&=\!\!&
\tilde{e}_{ik}^{}\otimes1+e^{\tilde{\sigma}_{11}^{}}\otimes
\tilde{e}_{ik}^{}
\end{eqnarray}
for $(ik)=(1\!-\!2)$, $(12)$, $(01)$ and
\begin{equation}
\Delta_{\xi_1^{}}(\tilde{e}_{1-1}^{})=\tilde{e}_{1-1}^{} \!
\otimes e^{-2\tilde{\sigma}_{11}^{}}+1\otimes
\tilde{e}_{1-1}^{}\!+\xi_1^{}\bigl(\tilde{e}_{12}^{}\wedge
\tilde{e}_{1-2}^{}+\tilde{e}_{01}^{}\otimes\tilde{e}_{01}^{}\bigr)
\bigl(e^{-\tilde{\sigma}_{11}^{}}\otimes
e^{-2\tilde{\sigma}_{11}^{}}\!\bigr). \label{jt10}
\end{equation}
The twisted Hopf structure of the subalgebra
$\widetilde{\mathfrak{osp}}_{2}(1|2):=w_{\xi_1^{}}
\mathfrak{osp}_{2}(1|2)w_{\xi_1}^{-1}\subset
\widetilde{\mathfrak{osp}}(1|4):=w_{\xi_1}{\mathfrak{osp}}(1|4)
w^{-1}_{\xi_1}$, generated by the elements
$\tilde{e}_{22}$, $\tilde{e}_{02}$, $\tilde{e}_{2-2}$,
$\tilde{e}_{-2-2}$, $\tilde{e}_{-20}$, is primitive, i.e.
\begin{eqnarray}
\Delta_{\xi_1^{}}(\tilde{e}_{ik}^{})\!\!&=\!\!&
\tilde{e}_{ik}^{}\otimes1+1\otimes \tilde{e}_{ik}^{} \label{jt14}
\end{eqnarray}
for $(ik)=(22)$, $(02)$, $(2\!-\!2)$, $(-2\!-\!2)$, $(-20)$. This
is not valid for the initial subalgebra $\mathfrak{osp}_{2}(1|2)$
what provides a main reason for the introduction of the similarity
transformation (\ref{jt7}) \cite{T3}. The formulas for coproducts
of the negative generator $\tilde{e}_{2-1}$, $\tilde{e}_{-2-1}$
and $\tilde{e}_{-10}$ have a more complicated form and we do not
give them here.

\underline{\it The second step of Jordanian type deformation}.
Since the subalgebra $\widetilde{\mathfrak{osp}}_{2}(1|2)$ is not
deformed, therefore, we can use the results of the paper
\cite{BLT}. Namely, we apply the twisting two-tensor
\begin{equation}
\begin{array}{rcl}
\tilde{F}_{22}^{}(\xi_2^{})\!\!&=\!\!&\bigl(w_{\xi_1^{}}\otimes
w_{\xi_1^{}}\bigr)\mathfrak{F}_{02}^{}(\xi_2^{})\,e^{e_{2-2}^{}
\otimes\sigma_{22}^{}}\bigl(w_{\xi_1}^{-1}\otimes
w_{\xi_1}^{-1}\bigr) =\tilde{\mathfrak{F}}_{02}^{}(\xi_2^{})\;
e^{\tilde{e}_{2-2}^{}\otimes\tilde{\sigma}_{22}^{}}
\end{array}\label{jt15}
\end{equation}
where
\begin{eqnarray}
\tilde{\mathfrak{F}}_{02}^{}\!\!&=&\!\!\sqrt{\frac{(1+
e^{\tilde{\sigma}_{22}})\otimes(1+e^{\tilde{\sigma}_{22}})}
{2(1+e^{\tilde{\sigma}_{22}}\otimes e^{\tilde{\sigma}_{22}})}}
\Bigl(1-2\xi_2^{}\frac{\tilde{e}_{02}^{}}
{1+e^{\tilde{\sigma}_{22}^{}}}\otimes\frac{\tilde{e}_{02}^{}}
{1+e^{\tilde{\sigma}_{22}^{}}}\Bigr)~, \label{jt16}
\\[5pt]
\quad\tilde{\sigma}_{22}^{}\!\!&=\!\!&
\frac{1}{2}\ln(1+\xi_2^{}\tilde{e}_{22}^{})\label{jt17}
\end{eqnarray}
to all generators of the $\xi_1^{}$-deformed  superalgebra
$\widetilde{\mathfrak{osp}}(1|4)$.

The twisted coproduct $\Delta_{\xi_1^{}\xi_2^{}}(\,\cdot\,):=
F_{22}^{}(\xi_2^{})\Delta_{\xi_1^{}}(\,\cdot\,)
F_{22}^{-1}(\xi_{2}^{})$ for the elements $\tilde{e}_{ik}$
belonging to Borel subalgebra of $\mathfrak{osp}(1|4)$ are given
by the formulas
\begin{eqnarray}
\Delta_{\xi_1^{}\xi_2^{}}
(e^{\pm\tilde{\sigma}_{11}^{}})\!\!&=\!\!&
e^{\pm\tilde{\sigma}_{11}^{}}\otimes
e^{\pm\tilde{\sigma}_{11}^{}}~, \label{jt18}
\\[7pt]
\Delta_{\xi_1^{}{}\xi_2^{}}(\tilde{e}_{12}^{})\!\!&=\!\!&
\tilde{e}_{12}^{}\otimes e^{\tilde{\sigma}_{22}^{}}+
e^{\tilde{\sigma}_{11}^{}}\otimes \tilde{e}_{12}^{}~, \label{jt19}
\end{eqnarray}\vskip-7pt
\begin{equation}
\begin{array}{rcl}
\Delta_{\xi_1^{}{}\xi_2^{}}(\tilde{e}_{01}^{})\!\!&= \!\!&
\tilde{e}_{01}^{}\otimes1+e^{\tilde{\sigma}_{11}^{}}\otimes
\tilde{e}_{01}^{}+\xi_2^{}\,\bigl(\tilde{e}_{12}^{}\otimes
\tilde{e}_{02}^{}-\tilde{e}_{02}^{}e^{\tilde{\sigma}_{11}^{}}
\otimes\tilde{e}_{12}^{}\bigr)\tilde{\Omega}\phantom{aaaaaaaaa}
\\[9pt]
&&-\xi_2^{2}\,\bigl(\tilde{e}_{12}^{}\tilde{e}_{02}^{}\otimes
\tilde{e}_{22}^{}+\tilde{e}_{22}^{}e^{\tilde{\sigma}_{11}^{}}
\otimes\tilde{e}_{12}^{}\tilde{e}_{02}^{}\bigr)
\bigl(\tilde{\omega}_{22}^{}\otimes
\tilde{\omega}_{22}^{}\bigr)\,\tilde{\Omega}~,
\end{array}
\label{jt20}
\end{equation}
\begin{equation}
\begin{array}{l}
\Delta_{\xi_1^{}{}\xi_2^{}}(\tilde{e}_{1-2}^{})=
\tilde{e}_{1-2}^{}\otimes e^{-\tilde{\sigma}_{22}^{}}+
e^{\tilde{\sigma}_{11}^{}}\otimes\tilde{e}_{1-2}^{}-
\xi_2^{}\tilde{e}_{2-2}^{}e^{\tilde{\sigma}_{11}^{}}\otimes
\tilde{e}_{12}^{}\,e^{-2\tilde{\sigma}_{22}^{}}
\\[9pt]
\phantom{aaaaaa}+\xi_2^{}\bigl(\tilde{e}_{01}^{}\otimes
\tilde{e}_{02}^{}e^{-\tilde{\sigma}_{22}^{}}+\tilde{e}_{02}^{}
e^{\tilde{\sigma}_{11}^{}}\otimes\tilde{e}_{01}^{}-
\xi_2^{}\tilde{e}_{12}^{}\tilde{e}_{02}^{}\tilde{\omega}_{22}^{}
e^{-\tilde{\sigma}_{22}^{}}\otimes\tilde{e}_{02}^{}
e^{-\tilde{\sigma}_{22}^{}}\qquad\quad
\\[9pt]
\phantom{aaaaaa}-\xi_2^{}\tilde{e}_{02}^{}e^{\tilde{\sigma}_{11}^{}}
\otimes\tilde{e}_{12}^{}\tilde{e}_{02}^{}\tilde{\omega}_{22}^{}
e^{-\tilde{\sigma}_{22}^{}}-\xi_2^{}\tilde{e}_{02}^{}
e^{\tilde{\sigma}_{11}^{}-\tilde{\sigma}_{22}^{}}\otimes\tilde{e}_{12}^{}
\tilde{e}_{02}^{}e^{-2\tilde{\sigma}_{22}^{}}\bigr)\tilde{\Omega}
\\[9pt]
\phantom{aaaaaa}+\xi_2^{2}\bigl(\tilde{e}_{01}^{}\tilde{e}_{02}^{}
\otimes\tilde{e}_{22}^{}e^{-\tilde{\sigma}_{22}^{}}-\tilde{e}_{22}^{}
e^{\tilde{\sigma}_{11}^{}}\otimes\tilde{e}_{01}^{}\tilde{e}_{02}^{}\bigr)
\bigl(\tilde{\omega}_{22}^{}\otimes\tilde{\omega}_{22}^{}\bigr)
\tilde{\Omega}
\\[9pt]
\phantom{aaaaaa}+\displaystyle{\frac{\xi_2^{2}}{2}}\bigl(\tilde{e}_{12}^{}
\otimes\tilde{e}_{22}^{}e^{-\tilde{\sigma}_{22}^{}}+
\tilde{e}_{22}^{}e^{\tilde{\sigma}_{11}^{}}\otimes\tilde{e}_{12}^{}
\bigr)\bigl(\tilde{\omega}_{22}^{}\otimes
\tilde{\omega}_{22}^{}\bigr)\tilde{\Omega}~,
\end{array}
\label{jt21}
\end{equation}\vskip-5pt
\begin{equation}
\begin{array}{rcl}
\Delta_{\xi_1^{}}(\tilde{e}_{1-1}^{})\!\!&=\!\!&\tilde{e}_{1-1}^{}
\!\otimes e^{-2\tilde{\sigma}_{11}^{}}+1\otimes\tilde{e}_{1-1}^{}
+\xi_1^{}\Bigl(\tilde{e}_{12}^{}\otimes\tilde{e}_{1-2}^{}
e^{\tilde{\sigma}_{22}^{}}-\tilde{e}_{1-2}^{}\otimes
\tilde{e}_{12}^{}e^{-\tilde{\sigma}_{22}^{}}
\\[9pt]
&&+\tilde{e}_{01}^{} \otimes\tilde{e}_{01}^{}+
\xi_2^{}\bigl(\tilde{e}_{01}^{}\otimes\tilde{e}_{02}^{}
\tilde{e}_{12}^{}\tilde{\omega}_{22}^{}e^{-\tilde{\sigma}_{22}^{}}-
\tilde{e}_{02}^{}\tilde{e}_{12}^{}\tilde{\omega}_{22}^{}
\otimes\tilde{e}_{01}^{}
\\[9pt]
&&+\tilde{e}_{01}^{}\tilde{e}_{02}^{}\tilde{\omega}_{22}^{}
\otimes\tilde{e}_{12}^{}e^{-\tilde{\sigma}_{22}^{}}+
\tilde{e}_{12}^{}\otimes\tilde{e}_{01}^{}\tilde{e}_{02}^{}
\tilde{\omega}_{22}^{}
\\[9pt]
&&-\displaystyle{\frac{1}{2}}
\tilde{e}_{12}^{}\tilde{\omega}_{22}^{}\otimes\tilde{e}_{12}^{}
e^{-\tilde{\sigma}_{22}^{}}-\displaystyle{\frac{1}{2}}
\tilde{e}_{12}^{}\otimes\tilde{e}_{12}^{}\tilde{\omega}_{22}^{}
-\tilde{e}_{2-2}^{}\tilde{e}_{12}^{}\otimes \tilde{e}_{12}^{}
e^{-\tilde{\sigma}_{22}^{}}
\\[9pt]
&&+\xi_2^{}\tilde{e}_{02}^{}\tilde{e}_{12}^{}\tilde{\omega}_{22}^{}
\otimes\tilde{e}_{02}^{}\tilde{e}_{12}^{}\tilde{\omega}_{22}^{}
e^{-\tilde{\sigma}_{22}^{}}\bigr)\Bigr)\bigl(e^{-\tilde{\sigma}_{11}^{}}
\otimes e^{-2\tilde{\sigma}_{11}^{}}\bigr)\;
\end{array}
\label{jt22}
\end{equation}
Here we use the notations:
$\xi_2^{}\tilde{e}_{22}^{}=e^{2\tilde{\sigma}_{22}^{}}-1$,
$\tilde{\omega}_{22}:=(e^{\tilde{\sigma}_{22}^{}}+1)^{-1}$,
$\tilde{\Omega}:=\Delta_{\xi_2}(\tilde{\omega}_{22}^{})=
(e^{\tilde{\sigma}_{22}^{}}\otimes
e^{\tilde{\sigma}_{22}^{}}+1)^{-1}$. The twisted Hopf structure
for the generators of subalgebra
$\widetilde{\mathfrak{osp}}_{2}(1|2)$ can be found in \cite{BLT}.

\section{Light-cone $\kappa$-deformation of
 the super-Poincar\'{e}\\ algebra ${\cal P}(3,1|1)$}
\setcounter{equation}{0}

In order to propose the application of deformed superalgebra
${\cal A} \in U_{\xi_1^{}\xi_2^{}}(\mathfrak{osp}(1|4)$) to the
descripton of anti-de-Sitter symmetries one should considered the
real forms which leave the classical $r$-matrices (\ref{cw11}),
(\ref{cw12}) skew-Hermitian , i.e.
\begin{equation}\label{phb1}
(r_1^{}(\xi_1^{}))^\star\,=\,-r_1^{}(\xi_1^{})~,
\qquad(r_2^{}(\xi_2^{}))^\star\,=\,-r_2^{}(\xi_2^{})~.
\end{equation}
where the $^\star$-conjugation is an antilinear
(super-)antiautomorphism and it defines the real form
$\mathfrak{sp}(4;\mathbb R)$ of $\mathfrak{sp}(4;\mathbb C)
\subset\mathfrak{osp}(1|4)$.

According to \cite{BLT} we consider two versions of the
$^\star$-conjugation which satisfy the condition (\ref{phb1}):

(i) The $^\dag$-conjugation is defined as follows
\begin{eqnarray}\label{phb2}
e^\dag_{jk}\,=\,-e_{jk}^{}~,\qquad e^\dag_{0j}\,=\,-i e_{0j}^{}
\qquad(j,k\,=\,\pm1,\pm2)~,
\end{eqnarray}
provided that
\begin{equation}\label{phb3}
(ab)^\dag\;= \;b^\dag a^\dag,\qquad (a\otimes
b)^\dag\;=\;(-1)^{\deg a\,\deg b}\,a^\dag\!\otimes b^\dag
\end{equation}
for any homogeneous elements $a,\,b\in U_{\xi_1^{}\xi_2^{}}
(\mathfrak{osp}(1|4))$.

(ii) The $^\ddag$-conjugation  we define by
\begin{eqnarray}\label{phb4}
e^\ddag_{jk}\,=\,-e_{jk}^{}~,\qquad e^\ddag_{0j}\,=\,-e_{0j}^{}
\qquad(i,j\,=\,\pm1,\pm2)~,
\end{eqnarray}
provided that
\begin{equation}\label{phb5}
(ab)^\ddag =(-1)^{\deg a\,\deg b} b^\ddag a^\ddag,\qquad (a\otimes
b)^\ddag =a^\ddag\!\otimes b^\ddag
\end{equation}
for any homogeneous elements $a,\,b\in U_{\xi_1^{}\xi_2^{}}
(\mathfrak{osp}(1|4))$. From the condition that $\sigma_{11}^{}$
(see (\ref{jt6})) and $\tilde\sigma_{22}^{}$ (see (\ref{jt17}))
are Hermitian, follows that the parameters $\xi_1^{}$, $\xi_2^{}$
are purely imaginary.

Let $M_{AB}$ ($A,B = 0,1,2,3,4$) describe the rotation generators
of the $AdS$ superalgebra
$\mathfrak{o}(3,2)\simeq\mathfrak{sp}(4;\mathbb R)$ with the
standard relations
\begin{eqnarray}\label{phb6}
\bigl[M_{AB},\, M_{CD}\bigr]\!\!&=\!\!&i\bigl(g_{BC}\,M_{AD}-
g_{BD}\,M_{AC}+g_{AD}\,M_{BC}-g_{AC}\,M_{BD}\bigr)~,
\\[7pt]\label{phb7}
M_{AB}\!\!&=\!\!&-M_{BA}~,\qquad M^\star_{AB}=M_{AB}
\end{eqnarray}
where $g_{AB} = \mathop{\rm diag}\,(1,-1,-1,-1,\;1)$. The
Cartan-Weyl (CW) generators $e_{jk}$ of $\mathfrak{sp}(4;\mathbb
R)$ can be realized in the terms of the generators $M_{AB}$ as
follows
\begin{equation}
\begin{array}{rcl}\label{phb8}
e_{1-2}^{}\!\!&=\!\!&i\bigl(M_{42}^{}+M_{21}^{}\bigr)~,\qquad\;
e_{12}^{}\;=\;i\bigl(M_{02}^{}+M_{32}^{}\bigr)~,\quad\;
e_{1-1}^{}\;=\;i\bigl(M_{03}^{}+M_{14}^{}\bigr)~,
\\[11pt]
e_{2-1}^{}\!\!&=\!\!&i\bigl(M_{42}^{}-M_{21}^{}\bigr)~,\quad\;
e_{-2-1}^{}\;=\;i\bigl(M_{02}^{}-M_{32}^{}\bigr)~,
\quad\; e_{2-2}^{}\;=\;i\bigl(M_{03}^{}-M_{14}^{}\bigr)~,
\\[11pt]
e_{11}^{}\!\!&=\!\!&i\bigl(M_{34}^{}+M_{04}^{}+M_{13}^{}-
M_{01}^{}\bigr)~,\quad
e_{-1-1}^{}\;=\;i\bigl(M_{34}^{}-M_{04}^{}-M_{13}^{}-
M_{01}^{}\bigr)~,
\\[11pt]
e_{22}^{}\!\!&=\!\!&i\bigl(M_{34}^{}+M_{04}^{}-M_{13}^{}+
M_{01}^{}\bigr)~,\quad
e_{-2-2}^{}\;=\;i\bigl(M_{34}^{}-M_{04}^{}+M_{13}^{}+
M_{01}^{}\bigr)~.
\end{array}
\end{equation}
Using these formulas one can write the boson (even) part of the
classical $r$-matrix (\ref{cw10}) in terms of the physical
generators $M_{AB}$\footnote{Compare with \cite{LLM2} where other
relations between the physical $AdS$ and CW bases were used.}:
\begin{equation}
\begin{array}{rcl}\label{phb9}
r_{\!1,2}^{b}(\xi_1^{},\xi_2^{})\!\!&=\!\!&\xi_1^{}
\Bigl({\displaystyle\frac{1}{2}}\;e_{1-1}^{}\wedge e_{11}^{}
+ e_{1-2}^{}\wedge e_{12}^{}\Bigr)+{\displaystyle\frac{1}{2}}
\xi_{2}^{}\,e_{2-2}^{}\wedge e_{22}^{}
\\[12pt]
\!\!&=\!\!&{\displaystyle\frac{1}{2}}(\xi_2^{}-\xi_1^{})
\Bigl(M_{14}^{}\wedge(M_{34}^{}+M_{04}^{})+M_{03}^{}
\wedge(M_{14}^{}-M_{01}^{})\Bigr)\,-
\\[9pt]
\!\!&-\!\!&{\displaystyle\frac{1}{2}}(\xi_1^{}+\xi_2^{})
\Bigl(M_{14}^{}\wedge(M_{13}^{}-M_{01}^{})+M_{03}^{}\wedge
(M_{34}^{}+M_{04}^{})\Bigr)\,-
\\[12pt]
\!\!&-\!\!&\xi_1(M_{42}^{}+M_{21}^{})\wedge(M_{02}^{}+M_{32}^{})~.
\end{array}
\end{equation}
There are two special case $\xi_1^{}=\xi_2^{}$ and
$\xi_1^{}=-\xi_2^{}$. We are interested in the first case
$\xi=\xi_1^{}=\xi_2^{}$:
\begin{equation}
\begin{array}{rcl}\label{phb10}
r^{b}(\xi):=r_{\!1,2}^{b}(\xi,\xi)
\!\!&=\!\!&-\xi\Bigl(M_{14}^{}\wedge(M_{13}^{}-M_{01}^{})+
M_{03}^{}\wedge(M_{34}^{}+M_{04}^{})\,+
\\[7pt]
&&\;\;+(M_{42}^{}+M_{21}^{})\wedge(M_{02}^{}+M_{32}^{}\Bigr)~.
\end{array}
\end{equation}
Introducing $\mu,\nu = 0,1,2,3$ and
\begin{equation}\label{phb11}
M_{\mu\,4}^{}\;=\;R\,{\cal P}_\mu^{}~,
\end{equation}
we obtain from (\ref{phb6}) the basic relation of $D=4$
algebra $AdS$:
\begin{equation}\label{phb12}
[{\cal P}_\mu, {\cal P}_\nu]=-\frac{1}{R^2}\,M_{\mu\nu}~.
\end{equation}
Using physical $AdS$ assignement of the generators $\{M_{AB}\}=
(M_j, N_j, {\cal P}_j, {\cal P}_0$), where $M_j=
\frac{1}{2}\epsilon_{jkl}M_{kl}$, $N_j=M_{0j}$ $(j,k,l=1,2,3)$ one
can write  the classical
$r$-matrix (\ref{phb10}) as follows:
\begin{equation}
\begin{array}{rcl}\label{phb13}
r^{b}(\xi)\!\!&=\!\!&\xi R\,\bigl({\cal P}_1^{}\wedge
(N_{1}^{}+M_{2}^{})+{\cal P}_{2}^{}\wedge(N_{2}^{}-M_{1}^{})+
{\cal P}_+^{}\wedge N_3^{}\bigr)
\\[9pt]
&&+\xi\,M_3^{}\wedge(N_2^{}-M_1^{})~,
\end{array}
\end{equation}
where ${\cal P}_{+}^{}={\cal P}_0^{}+{\cal P}_3^{}$.
Now we put $\xi=\frac{i}{\kappa R}$ and perform the limit
$R\to\infty$ (see (\ref{v3})). In such a way we obtain the
classical $r$-matrix  (\ref{v4}) describing light cone
$\kappa$-deformation of Poincar\'{e} algebra
($\lim\limits_{R\to \infty} {\cal P}_{\mu}^{}= P_\mu^{}$)
\begin{eqnarray}\label{phb14}
r^{b}_{\!\kappa}\!\!&
:=\!\!&\lim\limits_{R\to\infty}\,r^{b}\,\Bigl( \frac{i}{\kappa
R}\Bigr)=\frac{i}{\kappa}\,\Bigl(P_1^{}\wedge(N_1^{} +M_2^{})+
P_2^{}\wedge(N_2^{}-M_1^{})+P_+^{}\wedge N_3^{}\Bigr).
\end{eqnarray}
where the parameter $\kappa$ is real, and the Poincar\'{e}
algebra generators $(M_j, N_j, P_i, P_0)$ satisfy the relations
(\ref{v2}).

Similarly one can  discuss the classical ${\mathfrak{osp}}(1|4)$
$r$-matrix (\ref{cw10}) and its contraction limit. In order to
obtain finite result we put $\xi_1 = \xi_2$ and introduce in
accordance with (\ref{phb2}) and (\ref{phb4}) the real
${\mathfrak{osp}}(1|4)$ super-charges as follows
\begin{equation}\label{phb15}
e_{0\pm k}^{}\,=\,\sqrt{iR}\;Q_{\pm k}^{}
\qquad(k=\pm1,\pm2)~.
\end{equation}
One gets the formula for the super-Jordanian classical
${\mathfrak{osp}}(1|4)$ $r$-matrix
\begin{eqnarray}\label{phb16}
r(\xi)\!\!&:=\!\!&r_{\!12}^{}(\xi,\xi)\;=\;r^{b}_{\!12}(\xi,\xi)
-2iR\,\xi(Q_1^{}\wedge Q_1^{}+Q_2^{}\wedge Q_2^{})~.
\end{eqnarray}
which leads in the limit (\ref{v3})  to the following
super-Poincar\'{e} classical $r$-matrix:
\begin{eqnarray}\label{phb17}
r^{\mathrm susy}_{\!\kappa}\!\!&:=\!\!&\lim\limits_{R\to\infty}r
\Bigl(\frac{i}{\kappa R}\Bigr)\;=\;r^{b}_\kappa+\frac{2}
{\kappa}\,\bigl(Q_1^{}\wedge Q_1^{}+Q_2^{}\wedge Q_2^{}\bigr)~.
\end{eqnarray}
The classical $r$-matrix (\ref{phb17}) describes the superextension
of the light-cone $\kappa$-deformation of the Poincar\'{e} algebra.

In order to describe the adjoint action of ${\cal{P}}(3,1)$ on the
four real supercharges $Q_\alpha^{}$ $(\alpha=\pm1,\pm2)$) it is
convenient to introduce a $2$-component complex Weyl basis. In
terms of the Weyl basis $Q^{(\pm)}_1:=Q_1^{}\pm iQ_2^{}$,
$Q^{(\pm)}_2:=Q_{-1}^{}\pm iQ_{-2}^{}$ the commutation relations
read as follows
\begin{eqnarray}\label{phb18}
[M_j^{},Q^{(\pm)}_{\alpha}]\,=\,-\frac{i}{2}(\sigma_j^{})_
{\alpha\beta}^{}\,Q^{(\pm)}_{\beta},\quad
[N_j^{},Q^{(\pm)}_{\alpha}]\,=\,\mp\frac{i}{2}(\sigma_j^{})_
{\alpha\beta}^{}\,Q^{(\pm)}_{\beta},\quad
[P_\mu^{},Q^{(\pm)}_{\alpha}]\,=\,0,
\end{eqnarray}
and moreover
\begin{eqnarray}\label{phb19}
\{Q^{(\pm)}_{\alpha},\,Q^{(\pm)}_{\beta}\}\!\!&=\!\!&0~,\qquad
\{Q^{(+)}_{\alpha},\,Q^{(-)}_{\beta}]\;=\;2\bigl(
\delta_{\alpha\beta}^{}\,P_0^{}-
(\sigma_j^{})_{\alpha\beta}^{}\,P_j^{}\bigr)~,
\end{eqnarray}
where $\sigma_j^{}$ $(j=1,2,3)$ are $2\times2$ $\sigma$-matrices.
The spinor ${\bf Q}^{(+)}:=(Q^{(+)}_{1},Q^{(+)}_{2})$ transformes
as the left-regular representation and the spinor ${\bf
Q}^{(-)}:=(Q^{(-)}_{1},Q^{(-)}_{2})$ provides the right-regular
one.

Using the commutation relations (\ref{v2}) and (\ref{phb18}),
(\ref{phb19}) it easy to check that the $r$-matrix (\ref{phb17})
(and also (\ref{phb14})) is Jordanian type (\ref{v5}), (\ref{v6}),
where $h_{\gamma_0^{}}\rightarrow iN_3^{}$,
$e_{\gamma_0^{}}\rightarrow P_+^{}$, etc. Therefore we can
immediately read off the twisting two-tensor corresponding to this
$r$-matrix (see \cite{T3}) as an analog of the formulas
(\ref{jt2})--(\ref{jt6}). However we can obtain also a twisting
two-tensor corresponding to the classical $r$-matrix (\ref{phb17})
by applying the contraction $AdS$ limit to the full Jordanian type
twisting two-tensor of ${\mathfrak{osp}}(1|4)$. This full twist
$F(\xi_1^{},\xi_2^{})$ (\ref{jt1}) can be presented as follows
\begin{equation}
\begin{array}{rcl}
F(\xi_1^{},\xi_2^{})\!\!&=\!\!&\tilde{\mathfrak{F}}_{02}^{}
(\xi_2^{})\mathfrak{F}_{01}^{}(\xi_1^{})\Bigl
(\tilde{F}_{\sigma_{22}^{}}^{}\exp\bigl(\xi_1^{}e_{1-2}^{}\otimes
e_{12}^{}e^{-\sigma_{11}^{}}\bigr)\tilde{F}_{\sigma_{22}^{}}^{-1}
\Bigr)\tilde{F}_{\sigma_{22}^{}}^{}F_{\sigma_{11}^{}}^{}
\\[11pt]
\!\!&=\!\!&\tilde{\mathfrak{F}}_{02}^{}(\xi_2^{})
\mathfrak{F}_{01}^{}(\xi_1^{})\exp\bigl(\xi_1^{}e_{1-2}^{}\otimes
e_{12}^{}e^{-\sigma_{11}^{}\!-\tilde{\sigma}_{22}^{}}\bigr)
\tilde{F}_{\sigma_{22}^{}}^{}F_{\sigma_{11}^{}}^{}~.
\end{array}\label{phb20}
\end{equation}
Replacing here the mathematical generators $e_{jk}^{}$ by the
physical ones $M_{AB}$ and performing  the contraction limit we
obtain as a result the twisting two-tensor for the light-cone
$\kappa$-deformation of the super-Poincar\'{e} algebra
\begin{eqnarray}
F_{\kappa}^{}({\cal P}(3,1|1)\!\!&:=\!\!&\lim\limits_{R\to\infty}
F\Bigl(\frac{i}{\kappa R},\,\frac{i}{\kappa R}\Bigr)
\,=\,\mathfrak{F}_{\kappa}^{}(Q_2^{})\mathfrak{F}_{\kappa}^{}(Q_1^{})
F_{\kappa}^{}({\cal P}(3,1))~, \label{phb21}
\end{eqnarray}
where $F_{\kappa}^{}({\cal P}(3,1))$ is the twisting two-tensor of
the light-cone $\kappa$-deformation of the Poinca\-r\'{e} algebra
${\cal P}(3,1)$
\begin{eqnarray}
F_{\kappa}^{}({\cal P}(3,1))\!\!&:=\!\!&e^{\frac{i}{\kappa}\,
P_1^{}\otimes (N_1^{}+M_2^{})e^{-2\sigma_{\!+}^{}}}\,e^{\frac{i}
{\kappa}\,P_2^{}\otimes (N_2^{}-M_1^{})e^{-2\sigma_{\!+}^{}}}
\,e^{2iN_3\otimes\sigma_{\!+}^{}} \label{phb22}
\end{eqnarray}
and the super-factors $\mathfrak{F}_{\kappa}^{}(Q_\alpha)$
($\alpha=1,2$) are given by the formula
\begin{equation}
\mathfrak{F}_\kappa^{}(Q_\alpha^{})=\sqrt{\frac{(1+
e^{\sigma_{\!+}^{}})\otimes(1+e^{\sigma_{\!+}^{}})}
{2(1+e^{\sigma_{\!+}^{}}\!\otimes e^{\sigma_{+}^{}})}}
\biggl(1+\frac{2}{\kappa}\,\frac{Q_{\alpha}^{}}
{1+e^{\sigma_{\!+}^{}}}\otimes\frac{Q_{\alpha}^{}}
{1+e^{\sigma_{\!+}^{}}}\biggr)~,\label{phb23}
\end{equation}
\begin{equation}
\sigma_{\!+}^{}:=\frac{1}{2}\ln\Bigl(1+\frac{1}{\kappa}P_{+}^{}\Bigr)~.
\label{phb24}
\end{equation}
Since
$[N_1^{}+M_2^{},\sigma_{\!+}^{}]=[N_2^{}-M_1^{},\sigma_{\!+}^{}]=
[N_1^{}+M_2^{},N_2^{}-M_1^{}]=0$, all three exponentials on the
right side of (\ref{phb22}) mutually commute and they can be
written in any order. We add that the super-factors in
(\ref{phb21}) also mutually commute.

Using the twisting two-tensors (\ref{phb22}) and (\ref{phb23}) we
can calculate twisted coproducts and twisted antipodes for the
generators of the Poincar\'{e} and super-Poincar\'{e} algebras.
These formulas will be given in our future publication. It should
be noted that the twisting functions (\ref{jt1}), (\ref{phb22})
and (\ref{phb23}) satisfy the unitarity condition, i.e., for
example,
\begin{equation}
F_{\kappa}^{\star}({\cal P}(3,1))\;=\; F_{\kappa}^{-1}({\cal
P}(3,1))\bigr)~. \label{phb28}
\end{equation}
Therefore twisted coproduct
$\Delta_\kappa^{}(x):=F_\kappa^{}\Delta(x)F_\kappa^{-1}$ and
twisted antipode $S_\kappa^{}(x)$ are real under the
$^\star$-con\-jugation, i.e.
$\bigl(\Delta_\kappa(x)\bigr)^\star=\Delta_\kappa(x^\star)$,
$S_\kappa((S_\kappa(x^\star))^\star)=x$.

\section{Outlook}
In this paper we studied the Jordanian type deformation of
${\mathfrak{osp}}(1|4)$ with two deformation parameters $\xi_1$,
$\xi_2^{}$. If we interpret physically ${\mathfrak{osp}}(1|4)$ as
$D=4$ $AdS$ superalgebra, the parameters $\xi_1^{}$, $\xi_2^{}$
are dimensionless and the role of dimensionfull parameter takes
over $AdS$ radius $R$. The introduction of $D=4$
super-Poincar\'{e} limit requires the relation
$\xi_1^{}=\xi_2^{}=\xi$ and special contraction procedure
described by the formula (\ref{v3}) with $R$-dependent single
deformation parameter $\xi$. In such a way we obtain a new quantum
deformation of $D=4$ super-Poincar\'{e} algebra with $\kappa$ as
the deformation parameter. Recalling \cite{LNS} we can therefore
introduce by the contraction procedure two different
$\kappa$-deformations of $D=4$, $N=1$ supersymmetries:
\\
\renewcommand{\fboxsep}{0in}
\renewcommand{\fboxrule}{.1pt}
\newcommand{\rcurltag}{}
\unitlength=0.1cm {\begin{center}
{\begin{picture}(89.6,8.32)(7.36,117.12)
\put(-25.36,125.12){\makebox(0,.32)[lb]{Drinfeld-Jimbo}}
\put(25.84,123.44){\vector(1,0){43.4}}
\put(72.4,125.12){\makebox(0,0)[lb]{Standard $\kappa$-deformed D=4
}} \put(-25.36,118.64){\makebox(0,0)[lb]{deformation
$U_q(\mathfrak{osp}(1,4))$}}
\put(31.08,115.96){\makebox(0,0)[lb]{$\ln q=\frac{1}{\kappa
R}\;(R\to\infty)$}}
\put(72.4,118.64){\makebox(0,0)[lb]{Poincar\'{e} superalgebra
[15--17]}}
\end{picture}
\rcurltag }
\\[20pt]
\renewcommand{\fboxsep}{0in}
\renewcommand{\fboxrule}{.1pt}
\unitlength=0.1cm {\begin{picture}(89.6,8.32)(7.36,117.12)
\put(-25.36,124.12){\makebox(0,.32)[lb]{Jordanian type}}
\put(25.84,123.44){\vector(1,0){43.4}}
\put(72.4,124.12){\makebox(0,0)[lb]{Light-cone
$\kappa$-deformation}}
\put(-25.36,118.64){\makebox(0,0)[lb]{deformation
$U_{\xi_1,\xi_2}(\mathfrak{osp}(1|4))$}}
\put(27.08,115.96){\makebox(0,0)[lb]{$\xi_1=\xi_2=
\frac{i}{\kappa R}\;(R\to\infty)$}}
\put(72.4,118.64){\makebox(0,0)[lb]{of D=4 Poincar\'{e}
superalgebra}}
\end{picture}
\rcurltag }
\end{center}
It is known that light-cone $\kappa$-deformation of \hbox{$D=4$}
Poincar\'{e} algebra  with the classical $r$-matrix satisfying
CYBE can be extended to \hbox{$D=4$} conformal symmetries (see
e.g. \cite{LLM1}). Analogously, the light-cone
$\kappa$-deformation of $D=4$ Poincar\'{e} superalgebra can be
obtained by studying the suitable Jordanian type deformation of
\hbox{$D=4$} conformal superalgebra $\mathfrak{su}(2,2|1)$. The
Jordanian type deformations of $\mathfrak{su}(2,2|1)$ and in
particular the new embeddings of  $\kappa$-deformations of $D=4$
super-Poincar\'{e} algebra are now under consideration.

Finally we would like to mention that the $\kappa$-deformation of
$N$-extended $AdS$ supersymmetries can be described by the
Jordanian type deformations of $\mathfrak{osp}(N|4)$. An outline
of the mathematical framework describing the Jordanian type twist
quantization of $\mathfrak{osp}(M|2n)$ has been given recently in
\cite{T3}.

\subsection*{Acknowledgments}
The paper has been supported by KBN grant 1PO3B01828 (A.B.,J.L.) and
the grants RFBR-05-01-01086, INTAS-OPEN 03-51-3550
 (V.N.T.). The third author would like to thank
Institute for Theoretical Physics, University of Wroc{\l}aw
for hospitality and P.P.~Kulish for useful discussions.


\begin{thebibliography}{99}

\bibitem{LNRT}
 J. Lukierski, A. Nowicki, H. Ruegg and V.N. Tolstoy,
Phys. Lett. \textbf{264B}, 331 (1991).

\bibitem{OSWZ}
O. Ogievetsky, W.B. Schmidke, J. Wess and B. Zumino,
Commun. Math. Phys. \textbf{150}, 495 (1992).

\bibitem{D} V.K. Dobrev, Journ. Phys. \textbf{A26}, 1317 (1993).

\bibitem{M} S. Majid, Journ. Math. Phys. \textbf{34}, 2045 (1993);
\newline\url{http://xxx.lanl.gov/abs/hep-th/9210141}.

\bibitem
{MR} S. Majid, H. Ruegg, Phys. Lett. \textbf{B334}, 348 (1994);
\newline\url{http://xxx.lanl.gov/abs/hep-th/9405107}.

\bibitem 
{LRZ} J. Lukierski, H. Ruegg, W.J. Zakrzewski,  Ann. Phys.
\textbf{243}, 90 (1995);
\newline\url{http://xxx.lanl.gov/abs/hep-th/9312153}.

\bibitem 
{PW} P. Podle\'{s}, S.L. Woronowicz, Commun. Math. Phys.
\textbf{178}, 61 (1996);
\newline\url{http://xxx.lanl.gov/abs/hep-th/9412059}.

\bibitem 
{LMM} J. Lukierski, P. Minnaert, M. Mozrzymas, Phys. Lett.
\textbf{B371}, 215 (1996);
\newline\url{http://xxx.lanl.gov/abs/q-alg/9507005}.

\bibitem 
{KoM} P. Kosi\'{n}ski and P. Ma\'{s}lanka, in ``From Quantum Field
Theory to Quantum Groups", ed. B. Jancewicz, J. Sobczyk, World
Scientific, 1996, p. 41;
\newline\url{http://xxx.lanl.gov/abs/q-alg/9512018}.


\bibitem 
{BHOS} A. Ballesteros, F.J. Herranz, M.A. del Olmo, M. Santander,
Phys. Lett. \textbf{B351}, 137 (1995); \url{http://xxx.lanl.gov/abs/q-alg/9502019}.

\bibitem 
{H} F. Herranz, J. Phys. \textbf{A30}, 6123 (1997);
\url{http://arxiv.org/abs/q-alg/9704006}.

\bibitem 
{LLM1} J. Lukierski, V.D. Lyakhovsky and M. Mozrzymas, Phys. Lett.
\textbf{B538}, 375 (2002);
\newline\url{http://arxiv.org/abs/hep-th/0203182}.

\bibitem 
{LLM2} J. Lukierski, V.D. Lyakhovsky and M. Mozrzymas, Mod. Phys.
Lett. \textbf{18A}, 753 (2003);
\url{http://arxiv.org/abs/hep-th/0301056}.

\bibitem 
{KoLM2} P. Kosi\'{n}ski, J. Lukierski and P. Ma\'{s}lanka,
Phys. Rev. \textbf{D62}, 025004 (2000);
\newline\url{http://xxx.lanl.gov/abs/hep-th/9902037}.

\bibitem{M2}
M. Dimitrijevic, L. Jonke, L. Moller, E. Tsouchnika, J. Wess and M. Wohlgenannt,
 Eur. Phys. J. {\textbf C31}, 129 (2003);
\url{http://xxx.lanl.gov/abs/hep-th/0307149}.

\bibitem{R2}
A. Agostini, G. Amelino-Camelia and M. Arzano,
 Class. Quant. Grav. {\textbf 21}, 2179 (2004);
\url{http://xxx.lanl.gov/abs/gr-qc/0207003}.

\bibitem{M3}
M. Dimitrijevic, F. Meyer, L. Moller, and J. Wess,
 Eur. Phys. J. {\textbf C36}, 117 (2004);
\url{http://xxx.lanl.gov/abs/hep-th/0310116}.

\bibitem{M4}
M. Dimitrijevic, L. Jonke, L. Moller, E. Tsouchnika, J. Wess and M. Wohlgenannt,
 Czech. J. Phys. {\textbf 54}, 1243 (2004);
\url{http://xxx.lanl.gov/abs/hep-th/0407187}.

\bibitem{DFR1}
S. Doplicher, K. Fredenhagen and J.E. Roberts,
Phys. Lett. {\textbf B331}, 39 (1994); 
Commun. Math. Phys. {\textbf 172}, 187 (1995); 
\url{http://xxx.lanl.gov/abs/hep-th/0303037}.

\bibitem{ASS}
G. Amelino-Camelia, L. Smolin and A.Starodubtsev,
Class. Quant. Grav. 21 (2004) 3095; 
\url{http://xxx.lanl.gov/abs/hep-th/0306134}.

\bibitem{SWitten}
N. Seiberg and E. Witten, 
JHEP 9909, 032 (1999);
\newline\url{http://arxiv.org/abs/hep-th/9908142}.

\bibitem{FL} S. Ferrara and M. A. Lledo, 
JHEP 0005 (2000) 008;
 \newline\url{http://xxx.lanl.gov/abs/hep-th/0002084}.
 
\bibitem{KPT} D. Klemm, S. Penati and L. Tamassia, 
Class. Quant. Grav. 20 (2003) 2905; 
\newline\url{http://xxx.lanl.gov/abs/hep-th/0104190}.

\bibitem{BGN} J. de Boer, P. A. Grassi and P. van Nieuwenhuizen, 
Phys. Lett. B 574 (2003) 98; 
\newline\url{http://xxx.lanl.gov/abs/hep-th/0302078}.

\bibitem{Sei} N. Seiberg,
JHEP 0306, 010 (2003);
\url{http://arxiv.org/abs/hep-th/0305248}.

\bibitem 
{LNR} J. Lukierski, A. Nowicki,  H. Ruegg, in ``Topological and
Geometrical Methods in Field Theory", Eds. J. Mickelsson and O.
Pekonen, World Scientific, p. 202 (1992).
 
\bibitem 
{LNS} J. Lukierski, A. Nowicki, J. Sobczyk, J. Phys. \textbf{A26},
L1109 (1993).

 
\bibitem 
{KoLMS} P. Kosi\'{n}ski, J. Lukierski, P. Ma\'{s}lanka and J.
Sobczyk, J. Phys. \textbf{A28}, 2255 (1995);
\newline\url{http://xxx.lanl.gov/abs/hep-th/9405076}.

\bibitem 
{KoLM} P. Kosi\'{n}ski, J. Lukierski, P. Ma\'{s}lanka, in Proc. of
Nato Advanced Research Workshop "Noncomm. Struct. in Math. \&
Phys.", Kiev, 2000, Eds: S. Duplij and J. Wess, publ. Kluwer Acad.
Press, p. 79 (2001);
\url{http://xxx.lanl.gov/abs/hep-th/0011053}.

\bibitem{T3} V.N. Tolstoy,
in: Proc. of Internat. Workshop "Supersymmetries and Quantum
Symmetries (SQS'03)", Russia, Dubna, July, 2003, Eds: E. Ivanov
and A. Pashnev, publ. JINR, Dubna, p. 242 (2004);
\url{http://xxx.lanl.gov/abs/math.QA/0402433}.
 
\bibitem{CK1} E. Celegini and P.P. Kulish, J.
Phys. \textbf{A37}, no. 20, L211 (2004); Preprint POMI -- K6/2003.
\url{http://xxx.lanl.gov/abs/math.QA/0401272}.


\bibitem 
{KLM} P.P. Kulish, V.D. Lyakhovsky and A.I. Mudrov, J. Math. Phys.
\textbf{40}, 4569 (1999);
\newline\url{http://xxx.lanl.gov/abs/math.QA/9806014}.

\bibitem 
{LO} V.D. Lyakhovsky, M.A. del Olmo, Journ. Phys. \textbf{A32},
4541 (1999); \textbf{A32}, 5343 (1999);
\newline\url{http://xxx.lanl.gov/abs/math.QA/9903065}.

\bibitem{BLT}A. Borowiec, J. Lukierski and V.N. Tolstoy,
Mod. Phys. Lett. \textbf{18A}, 1157 (2003);
\newline\url{http://xxx.lanl.gov/abs/hep-th/0301033}.

\bibitem{CK2} E. Celegini, and P.P. Kulish,
J. Phys. \textbf{A31}, L79 (1998);
\newline\url{http://xxx.lanl.gov/abs/math.QA/9712024}.

\bibitem{ACS} N. Aizawa, R. Chakrabarti and J. Segal,
Mod. Phys. Lett. \textbf{18A}, 885 (2003);
\newline\url{http://xxx.lanl.gov/abs/hep-th/0301022}.

\bibitem{T1} V.N. Tolstoy, in Proc. "Group theoretical methods
in physics", Vol. I (Yurmala, 1985), ed. M.A. Markov, {\it VNU
Sci. Press, Utrecht}, p. 323 (1986).

\bibitem{O} O.V. Ogievetsky, Suppl. Rendic. Cir. Math. Palermo,
Serie II, No 37, p. 185 (1993); preprint MPI-Ph/92-99 (1992).
\end{thebibliography}
\end{document}